\documentclass[reprint, amsmath, amssymb, aps, superscriptaddress]{revtex4-1}
\usepackage{format}
\allowdisplaybreaks

\begin{document}

\preprint{APS/123-QED}

\title{ Thermoviscous Hydrodynamics in Non-Degenerate Dipolar Bose Gases }

\author{Reuben R.W. Wang}
\email{reuben.wang@colorado.edu}
\author{John L. Bohn}
\affiliation{JILA, University of Colorado, Boulder, Colorado 80309, USA}

\date{\today} 

\begin{abstract}

We present a hydrodynamic model of ultracold, but not yet quantum condensed, dipolar Bosonic gases. Such systems present both $s$-wave and dipolar scattering, the latter of which results in anisotropic transport tensors of thermal conductivity and viscosity. This work presents an analytic derivation of the viscosity tensor coefficients, utilizing the methods established in [Wang et al., arXiv:2205.10465], where the thermal conductivities were derived. 
Taken together, these transport tensors then permit a comprehensive description of hydrodynamics that is now embellished with dipolar anisotropy. An analysis of attenuation in linear waves illustrates the effect of this anisotropy in dipolar fluids, where we find a clear dependence on the dipole orientation relative to the direction of wave propagation.

\end{abstract}

\maketitle

\section{\label{sec:introduction} Introduction}

The field of ultracold dipolar physics has seen exciting progress in the last 2 decades \cite{Bohn17_Sci, Moses17_NatPhys, Liu21_FP}, brought about by technological advances in the cooling and trapping of magnetic lanthanide atoms \cite{Griesmaier05_PRL, Lu11_PRL, Aikawa12_PRL, Phelps20_arxiv, Chomaz22_arxiv}, and heteronuclear polar molecules \cite{Sage05_PRL, Ni08_Sci, Anderegg17_PRL, Voges20_PRL, Valtolina20_Nat}. Of note are recent experiments that have realized evaporative cooling in 3-dimensional polar molecular gases, made possible by electric field \cite{Li21_Nat} or microwave shielding \cite{Anderegg21_Sci, Schindewolf22_Nat}. The observed high ratio of elastic to inelastic collision rates permit long lived samples even at high densities, motivating study of dipolar induced anisotropic phenomena deep in the hydrodynamic regime. 
Ref.~\cite{Wang22_arxiv} takes the first step in formulating a hydrodynamic model of ultracold dipolar Bose gases, by deriving the transport tensor of thermal conductivity using the Chapman-Enskog procedure \cite{Uehling34_APS, Taxman58_APS, Monchick61_AIP, Chapman90_CUP}. Here we extend this formulation by deriving the viscosity tensor for gases consisting of dipolar  constituents. 

To construct these tensors, we assume molecular scattering cross sections due to ideal point dipoles aligned in a particular direction in space, as described in Ref.~\cite{Bohn14_PRA}. Fortunate for our analysis, the effective long-ranged molecular interaction potential between microwave shielded molecules is equivalent to the classical dipole potential \cite{Karman22_PRA, Schindewolf22_Nat}, permitting use of the collision cross sections derived in Ref.~\cite{Bohn14_PRA}. The same is apparently not strictly true for electric field shielded molecules \cite{Lassabliere22_arxiv}, but use of the same cross section has previously proven adequate in describing cross-dimensional thermalization experiments \cite{Li21_Nat, Wang21_PRA}.

These transport tensors then permit us to study the propagation of waves through the dipolar gas via the dispersion relation \cite{Sommerfeld60_APNY}, derived from the equations of conservation and constitution. Although extensively studied in quantum degenerate dipolar gases \cite{Wenzel18_PRL, Ticknor11_PRL, Block12_IOP, Andreev13_RPJ, Ronen10_PRA}, wave phenomena in their still thermal counterparts remain less treated. A goal of this work is, therefore, also to motivate deeper investigations of normal phase dipolar gases in the ultracold community. These systems promise a vast variety of dynamical phenomena still unexplored, for example anisotropies in flow behavior or in turbulence. 

The remainder of this manuscript is organized as follows: 
In Sec.~\ref{sec:anisotropic_viscosity}, we follow the procedure outlined in Ref.~\cite{Wang22_arxiv} to derive the transport tensor of viscosity from a microscopic theory of dipolar collisions. A linearization of the hydrodynamic equations allow us to extract the dispersion relation in Sec.~\ref{sec:attenuation}, where it is used to illustrate anisotropic wave attenuation. Finally, discussions and concluding remarks are drawn in sec.~\ref{sec:conclusions}.

\section{ \label{sec:anisotropic_viscosity} Anisotropic Viscosity }

\subsection{General}

\begin{figure}[ht]
    \centering
    \includegraphics[width=0.7\columnwidth]{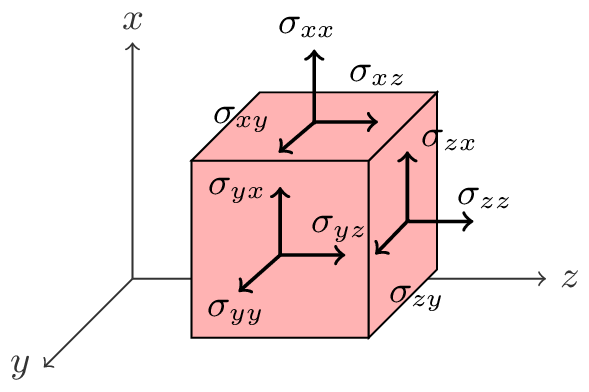} 
    \caption{ A visualization of the stresses (black arrows) on a differential fluid volume element (red cube) due to thermodynamic pressure and gradients in the velocity field. }
    \label{fig:viscous_stress_viz}
\end{figure}

Whereas a dilute gas is described in terms of the distribution of molecular velocities, a fluid in the hydrodynamic limit is described in terms of macroscopic observables such as the density $\rho$, velocity $\boldsymbol{U}$, and temperature $T$, all of which may vary in time and space.  The governing equations for these quantities, including in ultracold systems \cite{Nikuni98_JLTP}, are the equations of conservation \cite{Uehling33_APS}:
\begin{subequations} \label{eq:continuum_conservation_laws}
\begin{align}
    & \frac{ \partial \rho }{ \partial t } + \partial_j \left( { \rho U_j } \right) = 0, \\
    & \frac{\partial}{\partial t} \left( \rho U_i \right) + \partial_j \left( \rho U_j U_i \right) = \partial_j \sigma_{ij}, \\
    & \frac{\partial}{\partial t} (\rho T) + \partial_j \left( \rho T U_j \right) = \frac{ 2 m }{ 3 k_B } \left( \sigma_{ij} \partial_j U_i - \partial_j J_j \right), \label{eq:energy_conservation} 
\end{align}
\end{subequations}
where $\partial_i$ denotes the partial derivative with respect to coordinate $r_i$, and we invoke the summation conventions.  
The left-hand side of each of these equations denotes the rate of change of the relevant quantity at a given location in the fluid, including convective transport due to the fact that the fluid is in motion.  In the first, this derivative of the mass density $\rho$ is zero; this is the equation of continuity.  In the second equation, the change of the momentum density $\rho \boldsymbol{U}$ is given by forces acting in the fluid, thus giving the Navier-Stokes equation \cite{Navier23_MASIF, Stokes07_SEG}.  These forces are in turn given as gradients of the stress tensor $\sigma$, denoting the force on a surface with normal vector $\hat{ \boldsymbol{r}}_i$, in direction $\hat{ \boldsymbol{r}}_j$.  These forces are illustrated schematically in Figure \ref{fig:viscous_stress_viz}, where all vectors appear to have the same length, although this is not generally true in the fluid.  In the third equation of motion, for the mass-weighted temperature distribution, the rate of change depends on both the stress tensor and on the heat conduction vector $\boldsymbol{J}$.  

The equations (\ref{eq:continuum_conservation_laws}) are general in the absence of long-ranged forces between the molecules, and are described in Refs.~\cite{Bond65_AW, Chapman90_CUP, deGroot13_DP}. Additional external forces, such as the trap confining the atoms, can be added as necessary. To apply the equations of change to a particular fluid, such as the dipolar gas we have in mind, requires some constitutive relations describing the fluid.  For example, the stress tensor is written
\begin{align}
    \sigma_{ij} = -P \delta_{ij} + \tau_{ij}.
    \label{eq:stress_tensor_definition}
\end{align}
Here $P$ is the thermodynamic pressure, related to the density and temperature by the fluid's equation of state, which we will take as the ideal gas law in what follows. The remaining part,  $\boldsymbol{\tau}$, is the viscous stress tensor, i.e., the part arising from viscosity.  In a Newtonian fluid the viscous shear is assumed to be a linear function of the velocity gradients, so that generally, 
\begin{align} \label{eq:Newtonian_viscosity}
    & \tau_{ij} = \mu_{ijk\ell} \; \partial_\ell U_k,
\end{align} 
where $\boldsymbol{\mu}$ is in general a fourth-rank viscosity tensor and the flow velocity gradients $\partial_\ell U_k$, characterize the rate of strain on differential fluid volumes. 
We will see, however, that ultracold inelastic collisions leave only a symmeterized portion of $\partial_\ell U_k$ relevant. 
Meanwhile, the heat conduction vector is assumed to be linearly related to temperature gradients in accordance with Fourier's law,
\begin{align}
	J_i = - \kappa_{ij} \partial_i T,
	\label{eq:Fourier_law}
\end{align}
where $\boldsymbol{\kappa}$ is the thermal conductivity tensor,

The hydrodynamic relations of the dilute dipolar gas are therefore specified once the tensors $\boldsymbol{\mu}$ and $\boldsymbol{\kappa}$ are determined.  The thermal conductivity $\boldsymbol{\kappa}$ was previously derived in Ref.~\cite{Wang22_arxiv}. Here we turn our attention to the viscosity tensor $\boldsymbol{\mu}$.  

\subsection{Microscopic Theory}

The hydrodynamic variables are given by the velocity-averaged quantities
\begin{subequations}
\begin{align}
    & \rho(\boldsymbol{r}, t) = m n(\boldsymbol{r}, t) = \int d^3 v f(\boldsymbol{r}, \boldsymbol{v}, t) m, \\
    & \boldsymbol{U}(\boldsymbol{r}, t) = \frac{1}{n(\boldsymbol{r}, t)} \int d^3 v f(\boldsymbol{r}, \boldsymbol{v}, t) \boldsymbol{v}, \\
    & T(\boldsymbol{r}, t) = \frac{ 2 }{ 3 n(\boldsymbol{r}, t) k_B } \int d^3 v f(\boldsymbol{r}, \boldsymbol{v}, t) \frac{1}{2} m \boldsymbol{u}^2. 
\end{align}
\end{subequations}
where $f(\boldsymbol{r}, \boldsymbol{v}, t)$ denotes the phase space distribution of the molecules. The so-called peculiar velocity $\boldsymbol{u}(\boldsymbol{r}) = \boldsymbol{v} - \boldsymbol{U}(\boldsymbol{r})$, is the velocity of molecules relative to the local flow velocity. 

Similarly, the stress tensor is defined microscopically by the 
integral
\begin{align} 
    \sigma_{ij} &= -m \int d^3 u f(\boldsymbol{u})  u_i u_j.
\label{eq:sigma_def}
\end{align}
This integral computes the mean flux of momentum $m u_i$ through a surface with normal unit vector ${\hat r}_j$, thus describing a force per area on that surface.

In the spirit of the perturbative method of Chapman and Enskog, the phase space density is assumed to differ but little from its equilibrium value, 
\begin{align} \label{eq:close_to_equilibrium_ansatz}
    f \approx f_0 \left[ 1 + \Phi \right],
\end{align}
where $\Phi \ll 1$ and the equilibrium distribution of peculiar velocities is 
\begin{align} \label{eq:equilibrium_ansatz}
    f_0(\boldsymbol{u}, \beta) &= n_0(\beta) c_0(\boldsymbol{u}, \beta) \nonumber\\
    &= n_0(\beta) \left( \frac{ m \beta }{ 2 \pi } \right)^{3/2} \exp\left( - \frac{ m \beta }{ 2 } \boldsymbol{u}^2 \right),
\end{align}
$n_0$ is the gas equilibrium number density, $\beta = (k_B T)^{-1}$ and $\boldsymbol{u}^2 = u_k u_k$.

Reference \cite{Wang22_arxiv} argued, after a lengthy derivation based on \cite{Bond65_AW}, that a suitable variational {\it ansatz} for $\Phi$ is given in terms of gradients of temperature and velocity as 
\begin{align}
    \Phi = V_l b_{lk} \partial_k ( \ln T) + 2 m \beta W_{ij} a_{ijkl} D_{kl}.
    \label{eq:Phi_ansatz}
\end{align}
This expression is written in terms of the vector
\begin{align}
    & V_i  (\boldsymbol{u}) \equiv \left( \frac{  \beta m \boldsymbol{u}^2 }{ 2 } - \frac{ 5 }{ 2 } \right) u_i,
\end{align}
and symmetrized quantities
\begin{subequations} \label{eq:velocity_tensors}
\begin{align}
    & W_{ij}(\boldsymbol{u}) \equiv u_i u_j - \frac{ 1 }{ 3 } \delta_{ij} \boldsymbol{u}^2, \\
    & D_{ij}(\boldsymbol{U}) \equiv \frac{ 1 }{ 2 } \left( \partial_{j} U_i + \partial_i U_j \right) - \frac{ 1 }{ 3 } \delta_{ij} \partial_k U_k, \label{eq:flow_velocity_tensor}
\end{align}
\end{subequations}
where the coefficients $\boldsymbol{b}$ and $\boldsymbol{a}$ are to be determined variationally.  

The term in $\Phi$ involving $\ln T$ contributes to the thermal conductivity and was evaluated in Ref.~\cite{Wang22_arxiv}. Here we focus on the other term, in terms of which the stress tensor becomes 
\begin{align} 
    \sigma_{ij} &= -m \int d^3 u f_0(\boldsymbol{u}) \left[ 1 + \Phi(\boldsymbol{u}) \right] u_i u_j \\
    &=  -\frac{ n_0 }{ \beta } \delta_{ij} - 2 \left( m^2\beta \int d^3 u f_0(\boldsymbol{u}) u_i u_j W_{m n}  a_{m n k \ell}  \right) D_{k\ell}. \nonumber
    \label{eq:microscopic_stress}
\end{align}
This expression identifies the thermodynamic pressure as $P = n_0/\beta$, and the quantity in parentheses as related to the shear viscosity.  

\subsection{A Note on Symmetry}

The viscosity tensor as defined by Eq.~(\ref{eq:Newtonian_viscosity}) gives the stresses as linear combinations of the unsymmetrized second-rank tensor $\partial_l U_k$ giving the gradients of the fluid's  velocity components.  By contrast, the microscopic evaluation of stresses (\ref{eq:microscopic_stress}) from the Boltzmann equation, relates these stresses only to the symmetrized tensor $\boldsymbol{D}$.  The difference is telling: generally this tensor can be reduced in the usual way into the traceless, second-rank $\boldsymbol{D}$, along with an antisymmetric tensor $\boldsymbol{R}$ and a scalar,
\begin{align}
    \partial_{\ell} U_k = D_{k \ell} + R_{k \ell}  + \frac{1}{3} \delta_{k \ell} \grad\cdot\boldsymbol{U},
\end{align}
where
\begin{align}
    R_{k \ell} = \frac{ 1 }{ 2 } \left( \partial_{\ell} U_k - \partial_k U_{\ell} \right).
\end{align}
The absence of the antisymmetric tensor and the scalar from the expression connotes that there are no rotational viscosities, nor bulk viscosities in a dilute gas of particles with no internal degrees of freedom \cite{Chapman90_CUP, GadelHak95_JFE, Graves99_JTHT}. 
Without loss of generality, the viscous stress tensor can now be written in terms of just the traceless symmetric flow-velocity gradient tensor (a.k.a. strain rate tensor)
\begin{align}
    \tau_{i j} = 2 \mu_{ i j k \ell } D_{ k \ell }.
\end{align}
Note that this conclusion is independent of the form of the collision cross section of the molecules.

The relation between the two forms of the symmetrized tensors is conveniently handled via a contraction,
\begin{subequations}
\begin{align}
    & W_{ij}(\boldsymbol{u}) = {\cal I}_{i j k \ell} u_k u_{\ell} = u_k u_{\ell} {\cal I}_{k \ell i j} , \\
    & D_{ij}(\boldsymbol{U}) = {\cal I}_{i j k \ell} \partial_{\ell} U_k = \partial_{\ell} U_k {\cal I}_{k \ell i j},
\end{align}
\end{subequations}
with the traceless symmetric tensor
\begin{align}
    {\cal I}_{i j m n} = \frac{ \delta_{i m} \delta_{j n} + \delta_{i n} \delta_{j m} }{ 2 } - \frac{1}{3} \delta_{i j} \delta_{m n}.
\end{align}
Written in these terms, the expression for the shear stress tensor is  
\begin{align}
    \tau_{ij} = -2 m^2 \beta 
    \int d^3 u f_0 u_i u_j u_m u_n {\cal I}_{m n o p}  
    a_{o p k \ell} D_{ k \ell }. 
\end{align}
The integrand now consists of products of components of the peculiar velocity $\boldsymbol{u}$, multiplied by the known equilibrium velocity distribution.  All such integrals are readily evaluated (many leading to Kronecker delta functions), whereby the viscosity tensor ultimately becomes
\begin{align} \label{eq:viscosity_tensor_a}
    \mu_{ijk\ell} = -\frac{ 2 n_0 }{ \beta } {\cal I}_{i j m n} 
    a_{m n k \ell},
\end{align}
in terms of the variational parameters $\boldsymbol{a}$. These parameters must be determined by an approximate solution of the Boltzmann equation.

\subsection{Approximate Solution to the Boltzmann Equation \label{subsec:approximate_sol_BE} } 

To this end, we start with the Boltzmann equation for the local phase space distribution 
\begin{align} \label{eq:Boltzmann_equation}
    & \left( \frac{ \partial }{ \partial t } + v_i \partial_i \right) f(\boldsymbol{r}, \boldsymbol{v}, t) = {\cal C}[ f(\boldsymbol{r}, \boldsymbol{v}, t) ], 
\end{align}
where ${\cal C}[ f ]$
is the two-body collision integral \cite{Reif09_Waveland}. We ignore molecular finite-size effects required at higher densities, which would modify the collision integral above \cite{Chapman90_CUP, Hoffman65_PF}. At the current experimental regime of interest (detailed in Sec.~\ref{sec:attenuation}), we find that such effects only contribute $< 5\%$ changes to the transport coefficients.

Assuming the close-to-equilibrium phase space distribution of Eq.~(\ref{eq:close_to_equilibrium_ansatz}), we arrive at the {\it ansatz} in Eq.~(\ref{eq:Phi_ansatz}) and the approximate Boltzmann equation 
\begin{align} \label{eq:viscosity_BE}
    f_0 \: W_{k\ell} D_{k\ell} \approx 2 C[ f_0 W_{i j} ] a_{i j k \ell} D_{k\ell}.
\end{align}
Note that this refers to the portion of the simplified equation that pertains to viscosity, i.e., it does not include terms with temperature gradients.
Further details of this approximation are provided in Ref.~\cite{Wang22_arxiv}.

To obtain explicit forms for the variational coefficients $a_{i j k \ell}$ and hence the viscosities by Eq.~(\ref{eq:viscosity_tensor_a}), we rewrite the right-hand side  of  Eq.~(\ref{eq:viscosity_BE}) as
\begin{align}
    & 2 C[ f_0 W_{m n} ] a_{m n k \ell} D_{k \ell} \nonumber\\
    &\quad\quad = 2 C[ f_0 W_{m n} ] \left( {\cal I}_{m n r s} a_{r s k \ell} \right) D_{ k \ell } \nonumber\\
    &\quad\quad = -\frac{ \beta }{ n_0 } C[ f_0 W_{m n} ] \mu_{ m n k \ell } D_{ k \ell }.
\end{align}

Multiplying both sides of Eq.~(\ref{eq:viscosity_BE}) by $W_{i j}$ and integrating over $\boldsymbol{u}$ then gives
\begin{align} 
    & T_{i j k \ell} = M_{ i j m n } \mu_{ m n k \ell }. 
    \label{eq:integrated_Boltzmann}
\end{align}
where
\begin{subequations}
\begin{align} \label{eq:Enskog_collision_integrals}
    & T_{i j k \ell} = \int d^3 u f_0(\boldsymbol{u}) W_{i j} W_{k \ell} = \frac{ 2 n_0 }{ ( m \beta )^2 } {\cal I}_{i j k \ell}, \\
    & M_{ i j m n } = -\frac{ \beta }{ n_0 } \int d^3 u \: W_{i j} {\cal C}[ f_0 W_{m n} ]. \label{eq:M_collision_integrals}
\end{align}
\end{subequations}
The $M_{i j m n}$ integrals are particularly involved due to the highly anisotropic differential cross section of dipolar Bosons, for which the appropriately symmeterized scattering amplitude from Ref.~\cite{Bohn14_PRA} is given as
\begin{widetext}
\begin{align} \label{eq:scattering_amplitude}
    f_B\left(\hat{\boldsymbol{u}}^{\prime}_r, \hat{\boldsymbol{u}_r}\right)
    = 
    \frac{a_d}{\sqrt{2}} \left(
        \frac{ 4 }{ 3 } - \frac{ 2 a_s }{ a_d } - \frac{ 2 ( \hat{\boldsymbol{u}_r} \cdot \hat{\boldsymbol{{\cal E}}} )^{2} + 2 ( \hat{\boldsymbol{u}}^{\prime}_r \cdot \hat{\boldsymbol{{\cal E}}} )^{2} - 4 ( \hat{\boldsymbol{u}_r} \cdot \hat{\boldsymbol{{\cal E}}} ) ( \hat{\boldsymbol{u}}^{\prime}_r \cdot \hat{\boldsymbol{{\cal E}}} )( \hat{\boldsymbol{u}_r} \cdot \hat{\boldsymbol{u}}^{\prime}_r ) }{ 1 - (\hat{\boldsymbol{u}_r} \cdot \hat{\boldsymbol{u}}^{\prime}_r )^{2} }
    \right),
\end{align}
\end{widetext}
where $a_s$ is the $s$-wave scattering length, $\boldsymbol{u}_\mathrm{r} = \boldsymbol{u} - \boldsymbol{u}_1$ is the relative peculiar velocity between colliding molecules, primes denote post-collision velocities and $\hat{\boldsymbol{{\cal E}}}$ is the dipole alignment axis. This provides us the differential cross section via ${ d \sigma / d \Omega' } = \abs{ f_B \left( \hat{\boldsymbol{u}}^{\prime}_r, \hat{\boldsymbol{u}_r} \right) }^2$, with which to compute $\boldsymbol{M}$.  

After evaluating the integrals in Eq.~(\ref{eq:M_collision_integrals}), as detailed in App.~\ref{app:viscosity_integrals}, obtaining the $\boldsymbol{\mu}$ tensor now requires us to invert $\boldsymbol{M}$, which is most easily performed by converting $\boldsymbol{M}$ into its matrix representation denoted by an overhead circle, $\mathring{\boldsymbol{M}}$. This is done by mapping index pairs to single indices $(i, j) \rightarrow (i')$, via the relation
\begin{align} \label{eq:tensor2matrix_index}
    & i' = 3( j - 1 ) + i, 
\end{align}
rendering $M_{i j k \ell} \rightarrow \mathring{M}_{i' k'}$. In its $9 \times 9$ matrix representation, the inherent symmetries of $\boldsymbol{M}$ reduces its matrix rank from 9 to 5. This prevents us from inverting the matrix in its current representation, so we are now required to perform a change of basis transformation which decomposes the $9 \times 9$ matrix into a block-diagonal matrix with a $5 \times 5$ irreducible block. The desired change of basis matrix $\mathring{\boldsymbol{C}}$, is obtained by diagonalizing the isotropic tensor $\boldsymbol{{\cal I}}$ in its matrix representation, 
\begin{align}
    \mathring{\boldsymbol{{\cal I}}} = \mathring{\boldsymbol{C}} \left( \mathbb{I}_{5 \times 5} \oplus \boldsymbol{0}_{4 \times 4} \right) \mathring{\boldsymbol{C}}^{-1}, 
\end{align}
where $\mathbb{I}$ and $\boldsymbol{0}$ are the identity and zero matrices respectively, with dimensions specified by their subscripts, and $\oplus$ denotes a direct sum. Applying the transformation $\mathring{\boldsymbol{C}}$ to Eq.~(\ref{eq:integrated_Boltzmann}) gives
\begin{align}
    \mathring{\boldsymbol{C}}^{-1} \mathring{\boldsymbol{T}} \mathring{\boldsymbol{C}} 
    &=
    \mathring{\boldsymbol{C}}^{-1} \left( \mathring{\boldsymbol{M}} \mathring{\boldsymbol{\mu}} \right) \mathring{\boldsymbol{C}} \nonumber\\
    &= 
    \left( \mathring{\boldsymbol{C}}^{-1} \mathring{\boldsymbol{M}} \mathring{\boldsymbol{C}}^{-1} \right) \left( \mathring{\boldsymbol{C}} \mathring{\boldsymbol{\mu}} \mathring{\boldsymbol{C}} \right),
\end{align}
which leaves both sides of the equation above to only have a $5 \times 5$ non-trivial matrix block. The structure of these matrices are shown more explicitly by writing 
\begin{align}
    & \left[ \mathring{\boldsymbol{C}}^{-1} \mathring{\boldsymbol{T}} \mathring{\boldsymbol{C}} \right]_{5 \times 5} \oplus \boldsymbol{0}_{4 \times 4}
    = \nonumber\\ 
    &\quad\quad\quad 
    \left[ \left( \mathring{\boldsymbol{C}}^{-1} \mathring{\boldsymbol{M}} \mathring{\boldsymbol{C}} \right) \left( \mathring{\boldsymbol{C}}^{-1} \mathring{\boldsymbol{\mu}} \mathring{\boldsymbol{C}} \right)  \right]_{5 \times 5} \oplus \boldsymbol{0}_{4 \times 4}.
\end{align}
The direct sum with $\boldsymbol{0}_{4 \times 4}$ is trivial, so we can just consider the $5 \times 5$ irreducible subspace. This now allows us to effectively invert $\mathring{\boldsymbol{M}}$ by 
\begin{align}
    \left[ \mathring{\boldsymbol{C}}^{-1} \mathring{\boldsymbol{\mu}} \mathring{\boldsymbol{C}} \right]_{5 \times 5}
    &= 
    \left[
    \left( \mathring{\boldsymbol{C}}^{-1} \mathring{\boldsymbol{M}} \mathring{\boldsymbol{C}} \right)
    \right]_{5 \times 5}^{-1} 
    \left[
    \left( \mathring{\boldsymbol{C}}^{-1} \mathring{\boldsymbol{T}} \mathring{\boldsymbol{C}} \right)
    \right]_{5 \times 5},
\end{align}
and taking the direct sum of the expression above with $\boldsymbol{0}_{4 \times 4}$, to give 
\begin{align}
    \mathring{\boldsymbol{\mu}} 
    &=
    \mathring{\boldsymbol{C}}
    \left( \mathring{\boldsymbol{C}}^{-1} \mathring{\boldsymbol{M}}^+ \mathring{\boldsymbol{C}} \right)
    \left( \mathring{\boldsymbol{C}}^{-1} \mathring{\boldsymbol{T}} \mathring{\boldsymbol{C}} \right)
    \mathring{\boldsymbol{C}}^{-1} 
    = 
    \mathring{\boldsymbol{M}}^+ \mathring{\boldsymbol{T}},
\end{align}
where $\mathring{\boldsymbol{M}}^+$ is a pseudo-inverse of $\mathring{\boldsymbol{M}}$ defined by the procedure above, which satisfies $\mathring{\boldsymbol{M}}^+ \mathring{\boldsymbol{M}} \mathring{\boldsymbol{\mu}} = \mathring{\boldsymbol{\mu}}$. Finally, we apply the inverse mapping of Eq.~(\ref{eq:tensor2matrix_index}) to attain the rank-4 tensor form of $\boldsymbol{\mu}$. The explicit expressions for the viscosity coefficients are highly complicated, but tabulated in App.~\ref{app:viscosity_integrals} as a function of the dipole angle $\Theta$, defined to be the angle between $\hat{\boldsymbol{{\cal E}}}$, and $\hat{\boldsymbol{z}}$ in the $x,z$-plane.

\subsection{ Viscosities for Upright Dipoles }
 
In the case where the dipole and $\hat{\boldsymbol{z}}$ axes coincide (i.e. $\Theta = 0$), we find that the stress tensor simplifies greatly and can be written in a form more familiar in the usual theory of viscosity. 
That is, the stress tensor can be decomposed into a ``\textit{normal}" part which includes a proportionality with the symmetrized velocity gradients,
\begin{align}
    \tau^0_{ij} = 2 \mu^0_{ij} \circ 
    D_{i j}, 
\end{align}
and 2 additional ``\textit{anomalous}" stress terms which modify the radial diagonal stresses 
\footnote{ Such an anomalous stress also arises in nematic liquid crystals, where it is phenomenologically grouped together with the thermodynamic pressure to give an effective pressure \cite{DeGennes93_OUP}. }
\begin{align}
    \tau^{\prime}_{11} = \tau^{\prime}_{22} = 2 \mu' D_{3 3}.
\end{align}
The symbol $\circ$ denotes the Hadamard product (the element-wise product in which repeated indices remain unsummed).  The total stress is then written as the sum of the above two parts $\tau_{ij} = \tau^0_{ij} + \tau'_{ij}$, with the corresponding viscosity coefficients in this representation given explicitly as 
\begin{subequations} \label{eq:viscosities_Theta=0}
\begin{align}
    \mu^0_{13}  = \mu^0_{23} 
    &=
    \frac{ \mu_0 }{ 2 ( \mathscr{A}_2 - 3 \mathscr{A}_1 - {4 \tilde{a}_d^2 / 63} ) }, \\
    \mu^0_{12} = \mu^0_{11} = \mu^0_{22} 
    &=
    \frac{ \mu_0 }{ 2 \mathscr{A}_2 }, \\
    \mu^0_{33} 
    &=
    \frac{ \mu_0 }{ 2 \mathscr{A}_2 }
    +
    \frac{ 2 \mu_0 \mathscr{A}_1 }{ \mathscr{A}_2 ( \mathscr{A}_2 - 4 \mathscr{A}_1 ) }, \\
    \mu^{\prime} 
    &=
    \frac{1}{2} \left( \mu^0_{11} - \mu^0_{33} \right). 
\end{align}
\end{subequations}
Above, the $\boldsymbol{\mathscr{A}}$'s are polynomial functions of the reduced dipole length $\tilde{a}_d \equiv a_d / a_s$, in units of the scattering length
\begin{subequations}
\begin{align}
    & \mathscr{A}_1 = \frac{ 4 }{ 63 } \tilde{a}_d^2 - \frac{ 16 }{ 21 } \tilde{a}_d, \\
    & \mathscr{A}_2 = \frac{ 32 }{ 63 } \tilde{a}_d^2 - \frac{ 32 }{ 21 } \tilde{a}_d + 4. 
\end{align}
The dipole length is defined as $a_d = m d^2 / ( 8 \pi \epsilon_0 \hbar^2 )$ where $d$ is the electric dipole moment, $m$ is the molecular mass and $\epsilon_0$ is the vacuum permittivity. The viscosity coefficients are given in units of the Chapman-Enskog result
\end{subequations}
\begin{align}
    \mu_0 =  \frac{ 5 }{ 16 a_s^2 } \sqrt{ \frac{ m }{ \pi \beta } }.
\end{align}
corresponding to the viscosity of a gas of hard spheres with diameter $a_s$ \cite{Chapman90_CUP}.
 
We remark in passing that the stress tensor in this representation remains traceless, therefore the gas should not possess a bulk viscosity. This is a feature expected of monatomic gases in general.  It applies here since, at ultralow temperatures, only the ground state of the molecule is accessible.

In the limit where the scattering length remains finite and the dipole length goes to zero, we get $\mu^{\prime} = 0$ and all the other coefficients reduce to the same value,
\begin{align} \label{eq:zero_dipole_viscosity}
    \mu^0_{ij} (a_d = 0) 
    = 
    \frac{ \mu_0 }{ 8 }
    = 
    \frac{ 5 }{ 128 a_s^2 } \sqrt{ \frac{ m }{ \pi \beta } },
\end{align}
which is 8 times smaller than $\mu_0$ since the $s$-wave scattering cross section in Bose gases is $8 \pi a_s^2$ (instead of $\pi a_s^2$ for classical hard spheres), consistent with Ref.~\cite{Kavoulakis98_PRA}. 

\begin{figure}[ht]
    \centering
    \includegraphics[width=\columnwidth]{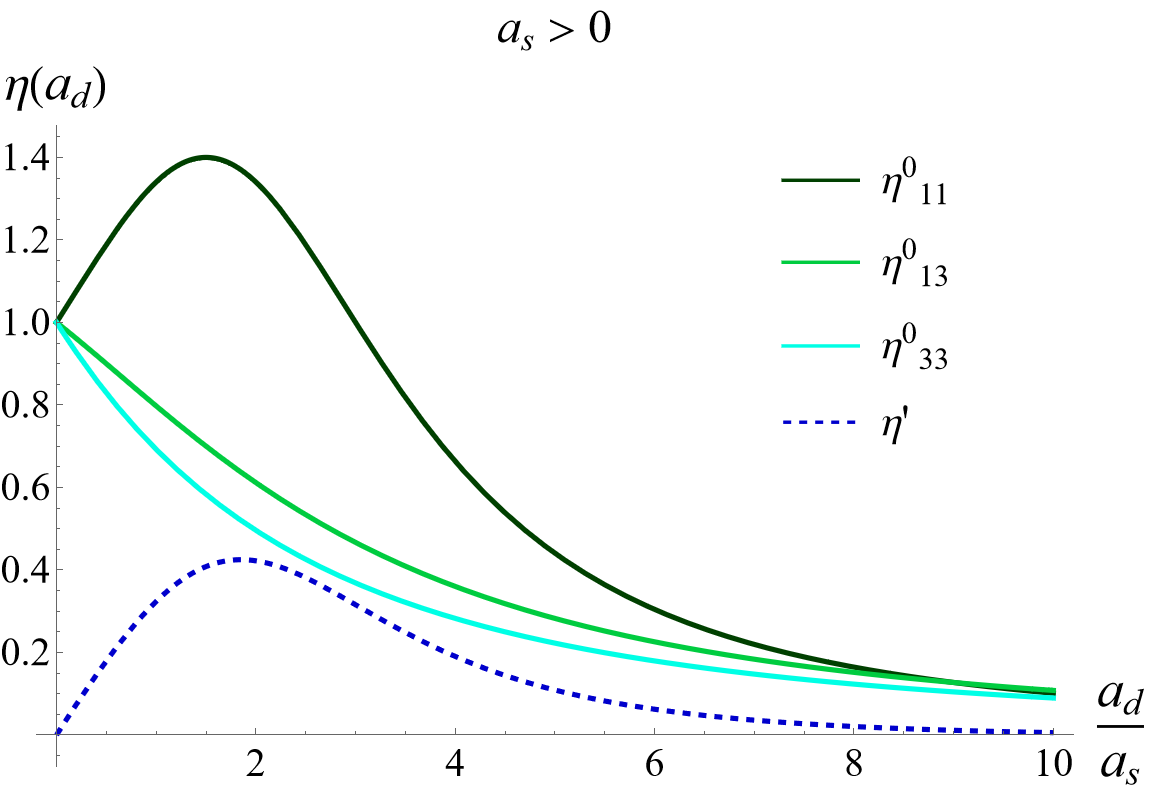}
    \caption{ The unique viscosity tensor elements with dipoles aligned along $\hat{\boldsymbol{z}}$, normalized by the finite scattering length isotropic viscosity of Eq.~(\ref{eq:zero_dipole_viscosity}), plotted as a function of $a_d / a_s$ from 0 to 10. The scattering length here is assumed positive ($a_s > 0$). }
    \label{fig:viscosities_vs_ad_pos_a}
\end{figure}

\begin{figure}[ht]
    \centering
    \includegraphics[width=\columnwidth]{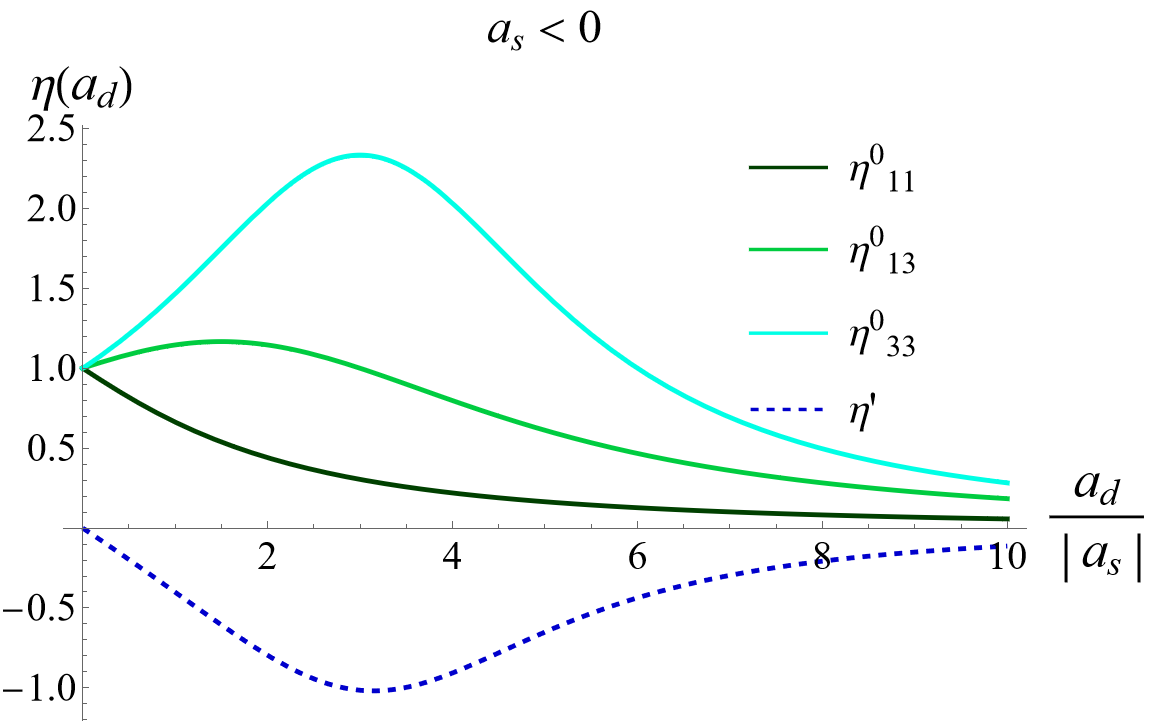}
    \caption{ The unique viscosity tensor elements with dipoles aligned along $\hat{\boldsymbol{z}}$, normalized by the finite scattering length isotropic viscosity of Eq.~(\ref{eq:zero_dipole_viscosity}), plotted as a function of $a_d / a_s$ from 0 to 10. The scattering length here is assumed negative ($a_s < 0$). }
    \label{fig:viscosities_vs_ad_neg_a}
\end{figure}

To illustrate the deviation from the isotropic result above as $a_d$ increases, we plot the the unique viscosities of Eq.~(\ref{eq:viscosities_Theta=0}) normalized by the isotropic viscosity of Eq.~(\ref{eq:zero_dipole_viscosity}):
\begin{align}
    \eta^0_{ij}(a_d) \equiv \frac{ \mu^0_{ij}(a_d) }{ \mu^0_{ij} (a_d = 0) }, \quad \eta'(a_d) \equiv \frac{ \mu'(a_d) }{ \mu^0_{ij} (a_d = 0) },
\end{align}
as a function of $\tilde{a}_d$ for both positive scattering length (Fig.~\ref{fig:viscosities_vs_ad_pos_a}) and negative scattering length (Fig.~\ref{fig:viscosities_vs_ad_neg_a}). To make these plots, we assume bosonic NaK molecules with a scattering length of magnitude $\abs{a_s} = 500 a_0$.

On the microscopic level,  the differential scattering of dipolar molecules exhibits a large anisotropy due to the dipole interaction itself, compounded by the interference of this scattering with the $s$-wave part characterized by a scattering length.  The way in which this interference plays out is via a tremendous variation of the relative viscosity coefficients as $\tilde{a}_d$ is varied.  Indeed, the ratio of $\mu^0_{33}$ to $\mu^0_{11}$ can vary from 0.36 to 9.3 based on the value and sign of $a_s$. Relative viscosities are therefore highly tunable in laboratory experiments via the scattering length with blue-detuned circularly polarized microwaves \cite{Lassabliere18_PRL} (also tunable in Lanthanide atoms via the multitude of Fano-Feshbach resonances \cite{Tang15_PRA, Patscheider21_PRA}), leading to phenomena yet to be explored.

Fig.~\ref{fig:viscosities_vs_ad_neg_a} reveals that the anomalous viscosity is negative when $a_s < 0$, implying a negative proportionality between radial viscous stresses and axial strain rates. Fortunately, this aberration does not imply an unwarranted dynamical instability since the positive axial viscous stress ensures the bulk fluid dilation remains identically zero.

\section{ \label{sec:attenuation} Attenuation in Dipolar Gases }

Having both the thermal conductivity and viscosity tensors now allows fluid dynamical studies in ultracold dipolar systems. As a first application, we analyze linear wave propagation through an initially uniform density gas. To do so, the density, flow velocity, and temperature variations are written in terms of fluctuations about their equilibrium distributions  
\begin{subequations} \label{eq:fluctuation_variables}
\begin{align}
    & \rho( \boldsymbol{r}, t ) = \rho_0 [ 1 + \chi( \boldsymbol{r}, t ) ], \\
    & U_i( \boldsymbol{r}, t ) = v_s \xi_i( \boldsymbol{r}, t ), \\
    & T( \boldsymbol{r}, t ) = T_0 [ 1 + \epsilon( \boldsymbol{r}, t ) ].
\end{align}
\end{subequations}
where $v_s = \sqrt{ { 5 / (3 m \beta_0) } }$ is the ideal gas speed of sound.
Plugging the form of Eq.~(\ref{eq:fluctuation_variables}) into Eq.~(\ref{eq:continuum_conservation_laws}) and assuming $\chi, \: \xi_i, \: \epsilon \ll 1$, allows a linearization to first-order in the fluctuations which gives 
\begin{subequations} \label{eq:linear_conservation_equations}
\begin{align}
    & \frac{ \partial \chi }{ \partial t } 
    \approx 
    - v_s { \partial_j \xi_j }, \\
    & \frac{ \partial \xi_i }{ \partial t } 
    \approx 
    - \frac{ 3 }{ 5 } v_s \partial_i ( \epsilon + \chi ) + \frac{ 3 \beta_0 v_s }{ 5 n_0 } \partial_j \tau_{ij}, \\ 
    & \frac{ \partial \epsilon }{ \partial t } 
    \approx 
    - \frac{ 2 }{ 3 } v_s \partial_j \xi_j + \frac{ 2 }{ 3 n_0 k_B } \kappa_{ij} \partial_i \partial_j \epsilon, \label{eq:linear_energy_conservation}
\end{align} 
\end{subequations}
as detailed in App.~\ref{app:linearized_balance_equations}. 

Solutions to Eqs.~(\ref{eq:linear_conservation_equations}) are obtained by utilizing the planewave ansatz for each dynamical variable, 
resulting in the system of equations
\begin{subequations}
\begin{align}
    & \omega \chi \approx v_s K_j \xi_j, \\
    & \omega \xi_i \approx \frac{ 3 }{ 5 } v_s K_i ( \epsilon + \chi ) - \frac{ i }{ \rho_0 } \mu_{i j k \ell} K_j K_{\ell} \xi_k, \\
    & \omega \epsilon \approx \frac{ 2 }{ 3 } v_s K_j \xi_j - \frac{ 2 i }{ 3 n_0 k_B } \kappa_{ij} K_i K_j \epsilon, 
\end{align}
\end{subequations}
Defining 
thermal conductivity and viscosity associated rates
\begin{subequations}
\begin{align}
    & \Gamma = - \frac{ 2 \kappa_{ij} }{ 3 n_0 k_B } K_i K_j, \\
    & \Lambda_{i k} = - \frac{ \mu_{i j k \ell} }{ \rho_0 } K_j K_{\ell},
\end{align}
\end{subequations} 
we get the linear system above written as the eigenvalue matrix equation
\begin{align} \label{eq:eigensystem}
    \begin{pmatrix}
        0 & v_s \boldsymbol{K}^T & 0 \\
        \frac{3}{5} v_s \boldsymbol{K} & i \boldsymbol{\Lambda} & \frac{3}{5} v_s \boldsymbol{K} \\
        0 & \frac{2}{3} v_s \boldsymbol{K}^T & i \Gamma
    \end{pmatrix}
    \begin{pmatrix}
        \chi \\
        \boldsymbol{\xi} \\
        \epsilon
    \end{pmatrix}
    =
    \omega \begin{pmatrix}
        \chi \\
        \boldsymbol{\xi} \\
        \epsilon
    \end{pmatrix}. 
\end{align} 

A dispersion relation is then obtained via the characteristic polynomial of Eq.~(\ref{eq:eigensystem}). Further analytical insight is gained by asserting only long wavelength ($\lambda$) excitations and large densities, which allow us to define the small parameter $\varepsilon = K_0 L$, where $K_0 = 2\pi / \lambda$ is the sourced-fixed wave-number, $L = ( \overline{\sigma} n_0 )^{-1}$ is the molecular mean-free path and $\overline{\sigma} = { 32\pi a_d^2 / 45 }$ is the angular averaged total cross section \cite{Bohn14_PRA}. The dispersion relation to first-order in $\varepsilon$ is then given as
\begin{align} \label{eq:dispersion_relation}
    & \omega ^3
    -
    i \omega^2 [ \Gamma + \tr(\boldsymbol{\Lambda}) ] 
    -
    \omega [ v_s K ]^2 \\
    &\quad +
    \frac{i}{5} v_s^2 
    \big[ 
    K_x^2 (3 \Gamma + 5 (\Lambda_{22}+\Lambda_{33})) \nonumber\\
    &\quad\quad\quad\:\:\: + K_y^2 (3 \Gamma + 5 (\Lambda_{11}+\Lambda_{33})) \nonumber\\
    &\quad\quad\quad\:\:\: + K_z^2 (3 \Gamma + 5 (\Lambda_{11}+\Lambda_{22})) \nonumber\\
    &\quad\quad\quad\:\:\: - 10 ( \Lambda_{12} K_y + \Lambda_{13} K_z ) K_x
    - 10 \Lambda_{23} K_y K_z  
    \big] 
    =
    0, \nonumber
\end{align}
where $\tr(\boldsymbol{\Lambda})$ denotes the trace of matrix $\boldsymbol{\Lambda}$.

To analyze the dispersion relation, we envision an experiment with a uniform density sample of ultracold $^{23}$Na$^{39}$K polar molecules. A wave generation source of constant frequency $\nu = \frac{ \omega }{ 2 \pi }$, is then applied that propagates waves along the $z$-direction. We focus our attention to the case where the scattering length is zero, which emphasizes the universal dipolar anisotropy while simplifying the viscosity and thermal conductivity expressions. We then solve the dispersion relation for the wave vector $K$ as functions of $\nu$ and $\Theta$, to first order in $\varepsilon$. This yields 2 pairs of $K$ solutions: 
\begin{widetext}
\begin{subequations}
\begin{align}
    & K_{1, \pm}(\omega, \Theta) 
    = 
    \pm \frac{ \omega }{ v_s }  \left[ 1 - \frac{ i 63 \omega }{ 16384 v_s a_d^2 n_0 } \sqrt{ \frac{ 5 }{ 3 \pi } } \left( 3 \cos (4 \Theta) - 21 \cos (2 \Theta) - 94 \right) \right], \\
    & K_{2, \pm}(\omega, \Theta)
    =
    \pm \sqrt{ \frac{ i \omega }{ v_s a_d^2 n_0 }  \sqrt{ \frac{ 5 }{ 3 \pi } } } 
    \left[
    \frac{ 
    16384 a_d^2 n_0 \sqrt{\pi } + i  ( 189 \cos (4 \Theta) - 1323 \cos (2 \Theta ) - 5922 ) \omega \sqrt{ m \beta_0 }
    }{
    128 \sqrt{ 105 \left( 6 \cos (4 \Theta) + 3 \cos (2 \Theta) + 7 \right) } 
    }
    \right].
\end{align}
\end{subequations}
\end{widetext}
The second solution pair $K_{2, \pm}$, has dominant imaginary terms which scale as $\sqrt{K_0 a_d^2 n_0} \sim K_0 / \sqrt{ K_0 L }$, causing corresponding wave solutions to attenuate within length scales of order $\sqrt{L}$. We therefore take it that these solutions do not support wave propagation. 

\begin{figure}[ht]
    \centering
    \includegraphics[width=\columnwidth]{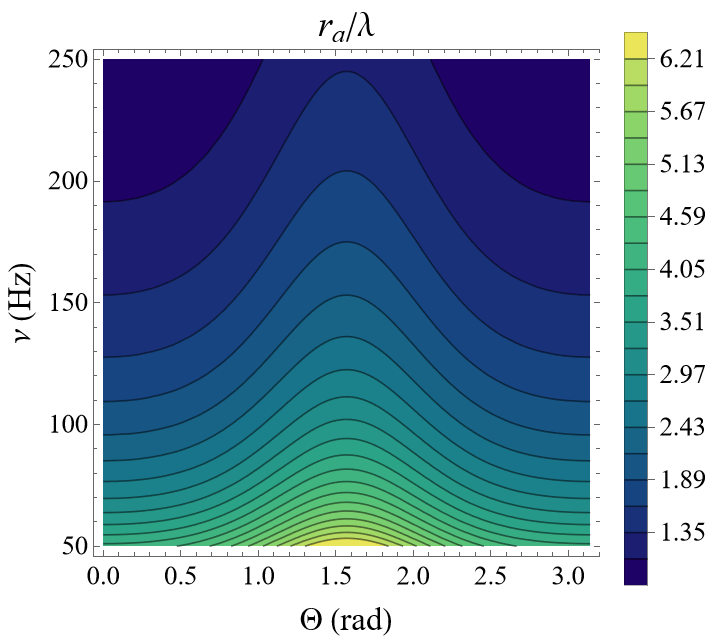}
    \caption{ The attenuation length $r_a$, normalized by the sourced-fixed wavelength $\lambda$, as a function of frequency $\nu$ (in Hertz, Hz) and dipole alignment angle $\Theta$ (in radians, rad).  }
    \label{fig:attenuation_length}
\end{figure}

As for $K_{1, \pm}$, these do support propagating waves with attenuation length $r_a$, set by
\begin{align}
    r_a &= \abs{ \Im[ K_{1, \pm} ] }^{-1},
\end{align}
and phase velocity 
\begin{align}
    v_p = \frac{ \omega }{ \abs{ \Re[ K_{1, \pm} ] } }.
\end{align} 
We see that to first-order in $\varepsilon$, the phase velocity is simply the frequency-independent ideal gas speed of sound $v_p = v_s$, whereas the attenuation length still retains a frequency dependence:
\begin{align}
    r_a(\omega) = \frac{1}{L} \left( \frac{ v_s }{ \omega } \right)^2 \frac{ 512 \sqrt{ { 15 / \pi } } }{ 7 ( 94 + 21 \cos(2 \Theta) - 3 \cos(4 \Theta)  ) }.
\end{align}
Therefore, we plot the ratio of attenuation length to the source-fixed wavelength $\lambda = \nu/v_s$ in Fig.~\ref{fig:attenuation_length} for $\Theta = 0$ to $\pi$ and $\nu = 50$Hz to $250$Hz. This frequency range is chosen to ensure $\varepsilon \approx 0.1$. 
The experimental parameters used to generate this plot are listed in Tab.~\ref{tab:system_parameter}, intended to reflect relevant experiments with NaK \cite{Voges20_PRL, Schindewolf22_Nat}. 

\begin{figure}[ht]
    \centering
    \includegraphics[width=\columnwidth]{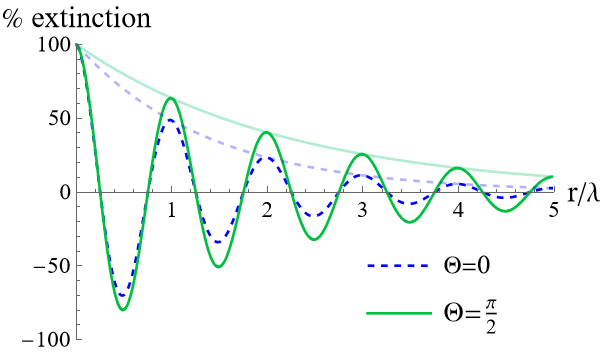}
    \caption{ The percentage extinction as a function of distance from the source $r$, normalized by the sourced-fixed wavelength $\lambda$, for $\Theta = 0$ (dashed blue curve) and $\Theta = \frac{\pi}{2}$ (solid green curve) with $\nu = 150$Hz. 
    The fainter curves are the decay envelopes associated to each extinction curve. }
    \label{fig:percentage_extinction}
\end{figure}

Fig.~\ref{fig:attenuation_length} showcases a clear variation of $r_a$ with the dipole angle $\Theta$, indicating that the attenuation of waves are highly dependent on the direction of propagation relative to the dipole orientation. In particular, waves that travel parallel to the dipole orientation attenuate faster in this case than those perpendicular to it. Moreover, the attenuation length is seen to decrease at higher frequencies as occurs with acoustic waves in ordinary dry air.  
The $\Theta$-dependence of wave attenuation is further made clear in Fig.~\ref{fig:percentage_extinction}, which plots the percentage extinction of the waveform $\cos( \abs{\Re[ K_{1} ]} r )\exp(-r / r_a) \times 100\%$, as a function of distance from the wave source $r$, for $\Theta = 0$ and $\frac{\pi}{2}$ with $\nu = 150$Hz. Fig.~\ref{fig:percentage_extinction} also plots the decay envelop with fainter colors for clarity. 

\begin{table}[ht]
\caption{\label{tab:system_parameter} 
Table of parameter values utilized to generate the plots for Bosonic $^{23}$Na$^{39}$K dipolar molecules. Da $= 1.661 \times 10^{-27}$ kg stands for Dalton (atomic mass unit), $a_{0} = 5.292 \times 10^{-11}$ m is the Bohr radius and D$ = 3.336 \times 10^{-30}$ Cm is a Debye. 
}
\begin{ruledtabular}
\begin{tabular}{l c c c}
    \multicolumn{1}{c}{\textrm{Parameter}} & \multicolumn{1}{c}{\textrm{Symbol}} & \multicolumn{1}{c}{\textrm{Value}} & \multicolumn{1}{c}{\textrm{Unit}} \\
    \colrule
    Relative molecular mass, & M$_\mathrm{r}$ & 62 & Da \\
    Effective electric dipole moment & $d_\mathrm{eff}$ & 0.75 & D \\
    Dipole length, & $a_d$ & $4.95 \times 10^4$ & $a_{0}$ \\
    Equilibrium number density, & $n_0$ & $5 \times 10^{12}$ & cm$^{-3}$  \\
    Equilibrium gas temperature, & $T_0$ & 250 & nK 
\end{tabular}
\end{ruledtabular}
\end{table}

We have thus far neglected long-range effects, which would modify the Navier-Stokes equation by adding the gradient of a mean-field potential
\begin{align}
    & U_\mathrm{mf}(\boldsymbol{r}) = \int d^3 r' n(\boldsymbol{r}') U_\mathrm{dd}(\boldsymbol{r} - \boldsymbol{r}'), \\
    \text{where}\quad & U_\mathrm{dd}(\boldsymbol{r}) = \frac{d^2}{ 4 \pi \epsilon_0 } \left( \frac{ 1 - 3 ( \hat{\boldsymbol{r}} \cdot \hat{\boldsymbol{{\cal E}}} ) }{ r^3 } \right)
\end{align}
is the potential between 2 point electric dipoles.
Close to uniform density with $\chi \approx 0.1$, the experimental parameters of Tab.~\ref{tab:system_parameter} yield $U_\mathrm{mf} / k_B T \approx 0.01$, validating this approximation.

As an added remark, we find that the viscosities of ultracold NaK molecules presented here are around $10^{11}$ times less than that of ordinary dry air at room temperature, $\mu_\mathrm{air}(T=300\text{K}) \approx 18.5 \: \mu$Pa$\cdot$s.

\section{ \label{sec:conclusions} Conclusions and Outlooks }

Recent experiments have demonstrated the ability to shield ultracold molecules from inelastic collisional losses, allowing long-lived dense samples of highly dipolar gases. As these molecular gases enter the hydrodynamic regime, a continuum model which includes the transport tensors of thermal conductivity and viscosity is warranted to describe the fluid dynamics. The thermal conductivities have been derived in Ref.~\cite{Wang22_arxiv}, and viscosities in this work, which now permit comprehensive phenomenological studies of Bosonic fluid systems that are enriched by dipolar anisotropy. 

As a first analysis, we looked at how the dipole orientation dependence of our derived transport tensors carry over into the attenuation of linear waves generated by a constant frequency source. We find that attenuation is most pronounced when the directions of wave propagation and dipole alignment coincide (i.e. $\Theta = 0$), whereas least attenuation occurs in the orthogonal configuration (i.e. $\Theta = \frac{\pi}{2}$). These results are illustrated with plots of the attenuation length and percentage extinction, that show a significant variation of these quantities with $\Theta$. Experiments which measure the attenuation length as a function of the dipole orientation could therefore provide an experimental means to extracting the anisotropic transport tensor coefficients.   

In the future, 
higher density corrections to the derived transport tensors can be included using the generalized Chapman-Enskog method \cite{Chapman90_CUP, Hoffman65_PF}. This modification would result in the emergence of a bulk viscosity and explorations of hydrodynamic phenomena not present in the current theory.  
At lower temperatures, the inclusion of quantum statistical effects \cite{Uehling33_APS} to our derived transport tensors could permit a normal-superfluid phase coupled hydrodynamic model for dipolar systems, extending the formalism established by Zaremba, Griffin and Nikuni \cite{Nikuni98_Spr, Zaremba99_Spr, Nikuni00_CJP}.
The mechanism to which dipolar  gases crossover from the dilute to hydrodynamic regimes might also be of interest to the ultracold community, such as in the context of evaporative cooling. Such a theory would interpolate the formulation presented in this work with that in Refs.~\cite{Wang20_PRA, Wang21_PRA}. 
Finally, the results of this work can also be extended to systems of Fermionic polar molecules and lanthanide atoms, by utilizing the scattering cross section for identical Fermions found in Ref.~\cite{Bohn14_PRA}.

\begin{acknowledgments}

This work is supported by the National Science Foundation under Grant Number PHY-2110327. 
The authors thank Eli Halperin for engaging discussions on the dispersion relations and their physical interpretations. 

\end{acknowledgments}

\appendix

\section{ \label{app:viscosity_integrals} Evaluation of the Viscosity Tensor }

To obtain the viscosity tensor, the collision integrals of Eq.~(\ref{eq:M_collision_integrals}) must be evaluated. These integrals are evaluated with methods identical to those described in Ref.~\cite{Wang22_arxiv}. However, the integrand differs slightly where instead of 
\begin{align}
    & N_{i k} \equiv \frac{ 2 m \beta }{ 5 n_0 } \int d^3 u \: V_i C[ f_0 V_k ], 
\end{align}
we are now evaluating
\begin{align}
    M_{ i j m n } = -\frac{ \beta }{ 2 n_0 } \int d^3 u \: W_{i j} {\cal C}[ f_0 W_{m n} ]. 
\end{align}
The collision integral is therefore similarly set-up by rewriting it in terms of center-of-mass (COM) and relative (r) velocity coordinates, $\boldsymbol{u}_\mathrm{COM}$ and $\boldsymbol{u}_\mathrm{r}$ respectively, which renders the product of equilibrium distributions as
\begin{align}
    & f_0(\boldsymbol{u})f_0(\boldsymbol{u}_1) = f_\mathrm{COM}(\boldsymbol{u}_\mathrm{COM}) f_\mathrm{r}(\boldsymbol{u}_\mathrm{r}). \label{eq:com_p_distribution}
\end{align}
Expanding the collision integral and writing it in terms of the COM and r coordinates leaves us with
\begin{align} \label{eq:Nik_collision_integral}
    M_{i j m n} &= -\frac{ \beta }{ 2 n_0 }
    \int d^3 {u}_\mathrm{COM} f_\mathrm{COM}(\boldsymbol{u}_\mathrm{COM}) \\
    &\quad\quad\:\:\: \times \int d^3 {u}_\mathrm{r} \: {u}_\mathrm{r} f_\mathrm{r}(\boldsymbol{u}_\mathrm{r}) W_{i j} \int d\Omega' \frac{ d \sigma }{ d \Omega' } \Delta W_{m n}. \nonumber 
\end{align}
The collision-varied quantity is written as
\begin{align}
    \Delta W_{i j} = \Delta ( u_i u_j ) &= \frac{1}{2} \left( u'_{r,i} u'_{r,j} - u_{r,i} u_{r,j} \right),
\end{align}
so integral over post-collision velocities with the differential cross section for dipolar Bosons given in Ref.~\cite{Bohn14_PRA}, becomes
\begin{align}
    I_{\Omega'} &\equiv \frac{1}{2} \int d\Omega'_r \frac{ d\sigma} { d\Omega'_r } \left( u'_{r,i} u'_{r,j} - u_{r,i} u_{r,j} \right).
\end{align}
We then utilize Mathematica \cite{Wolfram22} to evaluate and plug the above integral back into Eq.~(\ref{eq:Nik_collision_integral}) to obtain the elements of $M_{i j m n}$. 
Finally, we follow the procedure described in Sec.~\ref{sec:anisotropic_viscosity} to obtain the viscosity tensor elements. For brevity of presentation, we divide each viscosity by the isotropic viscosity as derived by Chapman and Enskog \cite{Chapman90_CUP}, denoted by a tilde: 
\begin{align}
    \tilde{\mu}_{i j k \ell} \equiv \mu_{i j k \ell} / \mu_0.
\end{align}
The 13 unit-free, unique and non-trivial viscosity tensor elements are tabulated below as a function of scattering length $a_s$, dipole length $a_d$ and dipole orientation angle $\Theta$:
\begin{widetext}
\begin{subequations} \label{eq:visocisty_coefficients}
\begin{align}
    & \tilde{\mu}_{1111} 
    =
    \frac{ 21 
    }{ 
    128 \mathscr{P}_1 \mathscr{P}_2 \mathscr{P}_3 } 
    \left[ 9 \mathscr{P}_1 \mathscr{P}_2 + 11 \mathscr{P}_1 \mathscr{P}_3 + 12 \mathscr{P}_2 \mathscr{P}_3 - 48 \mathscr{O}_1 \mathscr{P}_1 \cos (2 \Theta ) + 12 \mathscr{O}_3 \cos (4 \Theta ) \right],
    \\
    & \tilde{\mu}_{1113}
    =
    \frac{63 
    }{
    32 \mathscr{P}_1 \mathscr{P}_2 \mathscr{P}_3} 
    \left[ 2 \mathscr{O}_1 \mathscr{P}_1 \sin (2 \Theta ) - \mathscr{O}_3 \sin (4 \Theta ) \right], 
    \\
    & \tilde{\mu}_{1122}
    =
    -\frac{21
    }{
    8 \mathscr{P}_2 \mathscr{P}_3 } \left[ 1 + \mathscr{P}_1 - 3 \mathscr{O}_1 \cos (2 \Theta ) \right], 
    \\
    & \tilde{\mu}_{1133}
    =
    \frac{21 
    }{
    128 \mathscr{P}_1 \mathscr{P}_2 \mathscr{P}_3 } \left[ 3 \mathscr{P}_1 \mathscr{P}_2 - 7 \mathscr{P}_1 \mathscr{P}_3 - 12 \mathscr{P}_2 \mathscr{P}_3 - 12 \mathscr{O}_3 \cos (4 \Theta ) \right],
    \\
    & \tilde{\mu}_{1221} 
    =
    \frac{63 
    }{
    32 \mathscr{P}_1 \mathscr{P}_3 } \left[ \mathscr{P}_1 + \mathscr{P}_3 - 4 \mathscr{O}_2 \cos (2 \Theta ) \right], 
    \\
    & \tilde{\mu}_{1223}
    =
    \frac{ 63
    }{
    8 \mathscr{P}_1 \mathscr{P}_3 } \mathscr{O}_2 \sin (2 \Theta ),
    \\
    & \tilde{\mu}_{1331}
    =
    \frac{63
    }{
    128 \mathscr{P}_1 \mathscr{P}_2 \mathscr{P}_3 } \left[ \mathscr{P}_1 \mathscr{P}_2 + 3 \mathscr{P}_1 \mathscr{P}_3 + 4 \mathscr{P}_2 \mathscr{P}_3 - 4 \mathscr{O}_3 \cos (4 \Theta ) \right], 
    \\
    & \tilde{\mu}_{1322}
    =
    -\frac{63
    }{
    8 \mathscr{P}_2 \mathscr{P}_3 } \mathscr{O}_1 \sin (2 \Theta ), 
    \\
    & \tilde{\mu}_{1333}
    =
    \frac{63 
    }{
    16 \mathscr{P}_1 \mathscr{P}_2 \mathscr{P}_3 } 
    \left[ \mathscr{O}_1 \mathscr{P}_1 + \mathscr{O}_3 \cos (2 \Theta ) \right] \sin (2 \Theta ),
    \\
    & \tilde{\mu}_{2222}
    =
    \frac{ 21 }{ 16 \mathscr{P}_2 \mathscr{P}_3 } \left( 3 \mathscr{P}_2 + \mathscr{P}_3\right),
    \\
    & \tilde{\mu}_{2233}
    =
    -\frac{21 
    }{
    8 \mathscr{P}_2 \mathscr{P}_3 }
    \left[ 1 + \mathscr{P}_1 + 3 \mathscr{O}_1 \cos (2 \Theta ) \right], 
    \\
    & \tilde{\mu}_{2332}
    =
    \frac{63 
    }{
    32 \mathscr{P}_1 \mathscr{P}_3 }
    \left[ \mathscr{P}_1 + \mathscr{P}_3 + 4 \mathscr{O}_2 \cos (2 \Theta ) \right],
    \\
    & \tilde{\mu}_{3333}
    = 
    \frac{21 
    }{
    128 \mathscr{P}_1 \mathscr{P}_2 \mathscr{P}_3 } \left[ 9 \mathscr{P}_1 \mathscr{P}_2 + 11 \mathscr{P}_1 \mathscr{P}_3 + 12 \mathscr{P}_2 \mathscr{P}_3 + 48 \mathscr{O}_1 \mathscr{P}_1 \cos (2 \Theta ) + 12 \mathscr{O}_3 \cos (4 \Theta ) \right],
\end{align}
\end{subequations}
\end{widetext}
having defined the adimensonal polynomials of reduced dipole length $\tilde{a}_d = a_d / a_s$,
\begin{subequations}
\begin{align} 
    & \mathscr{O}_1 = \tilde{a}_d^2 - 12 \tilde{a}_d, \\
    & \mathscr{O}_2 = \tilde{a}_d^2 - 9 \tilde{a}_d, \\
    & \mathscr{O}_3 = 369 - 60 \tilde{a}_d - 4 \tilde{a}_d^2, \\
    & \mathscr{P}_1 = 63 + 12 \tilde{a}_d + 4 \tilde{a}_d^2, \\
    & \mathscr{P}_2 = 63 + 24 \tilde{a}_d + 4 \tilde{a}_d^2, \\
    & \mathscr{P}_3 = 63 - 24 \tilde{a}_d + 8 \tilde{a}_d^2. 
\end{align}
\end{subequations}
The latter 3 polynomials above that appear as denominators in the viscosities, $\mathscr{P}_1, \mathscr{P}_2$ and $\mathscr{P}_3$, do not have real roots for any combination of $a_s, a_d > 0$, preventing unphysical poles. 

Other non-trivial viscosity terms are specified by the tensor symmetry identities 
\begin{subequations} \label{eq:tensor_symmetries}
\begin{align}
    & \mu_{i j m n} = \mu_{j i m n}, \\
    & \mu_{i j m n} = \mu_{j i n m}, \\
    & \mu_{i j m n} = \mu_{m n i j}, \\
    & \mu_{i j m n} \delta_{i j} = 0, \\
    & \mu_{i j m n} \delta_{m n} = 0, \\
    & \mu_{i j m n} \delta_{i j m n} = 0,
\end{align}
\end{subequations}
where $\delta_{i j m n}$ is 1 if $i=j=k=\ell$ and $0$ otherwise. Repeated indices are summed over. All other unspecified tensor elements are zero.

\section{ \label{app:linearized_balance_equations} Linearizing the Balance Equations }

This section of the appendix details the linearization of the balance equations with the variational forms in Eq.~(\ref{eq:fluctuation_variables}). First, we shall take that the thermodynamic pressure to be given by the ideal gas law
\begin{align}
    & P = \frac{ n }{ \beta }. 
\end{align}
Then starting with the continuity equation, we have
\begin{align}
    & \frac{ \partial \rho }{ \partial t } + \partial_j \left( { \rho U_j } \right) = 0, \nonumber\\
    \Rightarrow\quad & 
    \rho_0 \frac{ \partial \chi }{ \partial t } + \rho_0 v_s \left( { \partial_j \xi_j } + { \xi_j \partial_j \chi } \right) = 0, \nonumber\\
    \Rightarrow\quad & 
    \frac{ \partial \chi }{ \partial t } + \sqrt{ \frac{ 5 }{ 3 m \beta_0 } } { \partial_j \xi_j } \approx 0.
\end{align}
The Navier-Stokes equation is then 
\begin{align}
    & \frac{\partial}{\partial t} \left( \rho U_i \right) + \partial_j \left( \rho U_j U_i \right) = - \partial_i P + \partial_j \tau_{ij}, \nonumber\\
    \Rightarrow\quad & 
    v_s \rho_0 \left( ( 1 + \chi ) \frac{ \partial \xi_i }{ \partial t } + \xi_i \frac{ \partial \chi }{ \partial t } \right) \nonumber\\
    &\quad + v_s^2 \rho_0 \Big( \partial_j (\xi_j \xi_i) + \xi_j \xi_i \partial_j \chi \Big) \nonumber\\
    &\quad = - \frac{ n_0 }{ \beta_0 } ( 1 + \epsilon ) \partial_i \chi - \frac{ n_0 }{ \beta_0 } (1 + \chi) \partial_i \epsilon
    + \partial_j \tau_{ij}, \nonumber\\
    \Rightarrow\quad & 
    \frac{ \partial \xi_i }{ \partial t } \approx - \sqrt{ \frac{ 3 }{ 5 m \beta_0 } } \partial_i ( \epsilon + \chi ) + \frac{ 1 }{ n_0 } \sqrt{ \frac{ 3 \beta_0 }{ 5 m } } \partial_j \tau_{ij}. r_i,
\end{align}
Finally, we have the energy balance equation as
\begin{align}
    & \frac{\partial}{\partial t} (\rho T) + \partial_j \left( \rho T U_j \right) 
    =
    \frac{2 m}{3 k_B} ( \sigma_{ij} \partial_j U_i - \partial_j J_j ), \nonumber\\
    \Rightarrow\quad & 
    \rho_0 T_0 \left( (1 + \chi) \frac{ \partial \epsilon }{ \partial t } + ( 1 + \epsilon ) \frac{ \partial \chi }{ \partial t } \right) \nonumber\\
    &\quad + v_s \rho_0 T_0 \Big[ (1 + \chi) (1 + \epsilon) \partial_j \xi_j \nonumber\\
    &\quad\quad\quad + ( 1 + \chi ) \xi_j \partial_j \epsilon + (1 + \epsilon) \xi_j \partial_j \chi \Big] \nonumber\\
    &\quad 
    = 
    \frac{2 m}{3 k_B} ( v_s \sigma_{ij} \partial_j \xi_i - \partial_j J_j), \nonumber\\ 
    \Rightarrow\quad & 
    \frac{ \partial }{ \partial t } \left( \epsilon + \chi \right) + \frac{ 5 }{ 3 } \sqrt{ \frac{ 5 }{ 3 m \beta_0 } } \partial_j \xi_j \approx -\frac{ 2 }{ 3 } \frac{ \beta_0 }{ n_0 } \partial_j J_j, \nonumber\\
    \Rightarrow\quad & 
    \frac{ \partial \epsilon }{ \partial t } + \frac{ 2 }{ 3 } \sqrt{ \frac{ 5 }{ 3 m \beta_0 } } \partial_j \xi_j \approx -\frac{ 2 }{ 3 } \frac{ \beta_0 }{ n_0 } \partial_j J_j.
\end{align}
In summary these grant us the closed set of equations in Eq.~(\ref{eq:linear_conservation_equations}) of the main text. 

\nocite{*}

\bibliography{main.bib} 

\end{document}